\documentclass[twocolumn,twoside,slac_two]{revtex4}

\usepackage{graphicx}

\usepackage{fancyhdr}
\newlength{\twocolfigwidth}
\newlength{\onecolfigwidth}
\newlength{\twothirdscolfigwidth}
\newcommand{\gammaRays}{$\gamma$ rays}
\newcommand{\gammaRayHyph}{$\gamma$-ray}

\newcommand{\irf}[1]{\texttt{#1}}

\newcommand{\Fermi}{{\textit{Fermi}}}
\newcommand{\fermi}{\Fermi}

\begin{document}

\title{\fermi-LAT data reprocessed with updated calibration constants}

\author{J. Bregeon}
\affiliation{INFN-Sezione di Pisa, Largo Pontecorvo 3, Pisa, I-56127, Italy}

\author{E. Charles, M. Wood}
\affiliation{KIPAC and SLAC National Accelerator Laboratory, Menlo Park, CA, USA}

\author{ for the \fermi-LAT collaboration}

\begin{abstract}
    Four years into the mission, the understanding of the performance of the
    \Fermi\ Large Area Telescope (LAT) and data analysis have increased
    enormously since launch. Thanks to a careful analysis of flight data, we
    were able to trace back some of the most significant sources of systematic
    uncertainties to using non-optimal calibration constants for some of the
    detectors.
    In this paper we report on a major effort within the LAT Collaboration to
    update these constants, to use them to reprocess the first four~years of
    raw data, and to investigate the improvements observed for low- and
    high-level analysis. The Pass~7 reprocessed data, also known as
    \irf{P7REP} data, are still being validated against the original Pass~7
    (\irf{P7}) data by the LAT Collaboration and should be made public, along
    with the corresponding instrument response functions, in the spring of 2013.
\end{abstract}

\maketitle

\thispagestyle{fancy}

\section{\Fermi-LAT calibration constants}

The \Fermi-LAT data acquisition system electronics relies on a number of
calibration constants (we refer the reader to~\citet{REF.2009.OnOrbitCalib} for
more details). Most of them are either stable or drift very slowly ($\sim 1\%$
per year). We keep track of the calibration constants for a definite time span
with a dedicated database.

For the Anticoincidence Detector (ACD) subsystem, pedestals, low- and high-range
gains need to be calibrated. For the Tracker (TKR) subsystem, hot and dead
strips have to be identified and the time-over-threshold charge scale must be
defined. For the Calorimeter (CAL) subsystem, pedestals, gains, electronics
non-linearity and cross-talk are measured through periodic triggers and charge
injection runs. In addition, two intrinsic characteristics of the CAL CsI(Tl)
crystals must be calibrated: the light yield and the light tapering. We note
in passing that other calibrations, such as the alignment of the TKR, appear to
have changed negligibly over the mission to date.

The calibrations constants used for the \irf{P7REP} data caused two
significant changes with respect to the \irf{P7} data: a slight shift in the
LAT energy scale, and an improvement the shower imaging resolution of the CAL.

\subsection{Energy scale}

\begin{figure}[htbp!]
  \includegraphics[width=\linewidth]{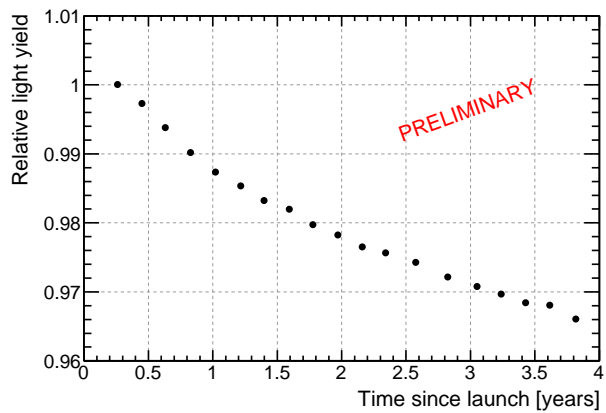}
  \caption{Relative variation of the absolute energy scale, as measured from
    the pathlength-corrected energy deposition of on-orbit minimum ionizing
    protons, throughout the first four years of the mission.}
  \label{fig:cal_light_yield}
\end{figure}

On-orbit, we measure the CsI(Tl) light yield by selecting minimum ionizing
protons from the LAT triggers. The calorimeter CsI(Tl) crystals suffer
radiation damage, which induces a decrease of the scintillation efficiency by
$\sim 1\%$ per year as shown in figure~\ref{fig:cal_light_yield}.
\irf{P7REP} data benefited from up-to-date calibration constants for the
CsI(Tl) light yield.

\subsection{Light asymmetry}

The attenuation of light along the longitudinal axis of each CsI(Tl) crystal
has to be calibrated on a regular basis as the light asymmetry between the two
crystal ends is used to reconstruct the longitudinal position of the energy
deposit. The calibration is performed using cosmic-ray heavy ions that release
their energy only via ionization: heavy ions provide well localized high-energy
depositions, very suitable for this purpose. 
As we have learned since launch, the measurement of light asymmetry has a
direct and significant impact on the determination of the energy centroid in
the calorimeter, which is used in the the tracking stage of the event
reconstruction. This, in turn, determines the instrument point-spread function
(PSF). \irf{P7REP} data were processed with light tapering calibration updated
every 2 months.

\begin{figure*}[t]
  \includegraphics[width=0.5\linewidth]{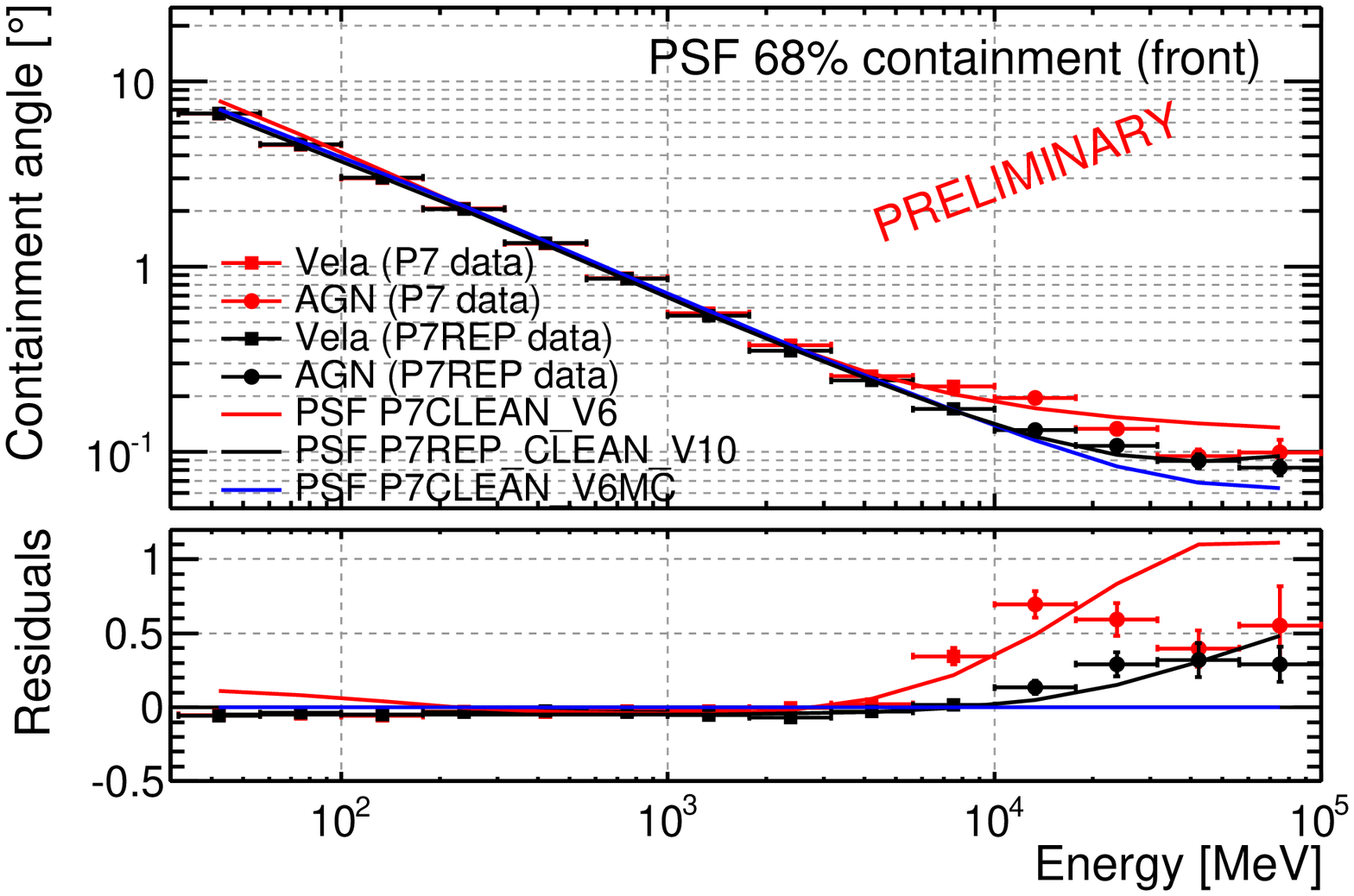}%
  \includegraphics[width=0.5\linewidth]{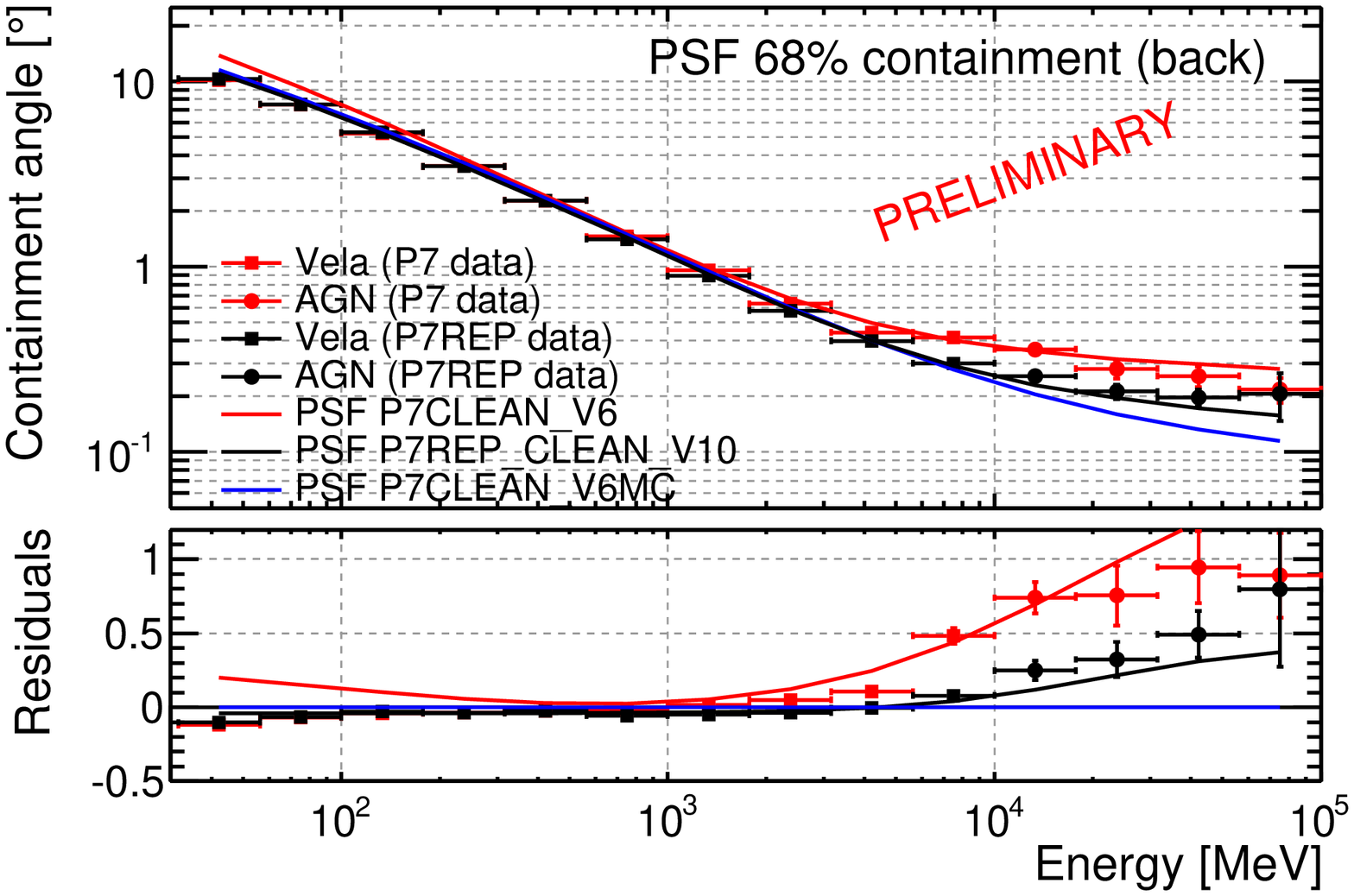}
  \caption{PSF 68\% containment radii for front- and back-converting events,
    derived from on-orbit data (using the Vela pulsar and a stacked sample of
    AGN) for both \irf{P7CLEAN} and \irf{P7REP\_CLEAN}.
    The solid lines show the PSF parameterizations for the \irf{P7CLEAN\_V6MC},
    \irf{P7CLEAN\_V6} and \irf{P7REP\_CLEAN\_V10} sets of IRFs.}
  \label{fig:psf68}
\end{figure*}

\section{Instrument Response Functions}

The analysis of \fermi-LAT data requires a number of tools and ancillary data
products including templates for the isotropic and Galactic diffuse emission
and instrument response functions (IRFs). All components will be updated to
match the characteristics of \irf{P7REP} data and will be released together
with the \irf{P7REP} data for the purpose of improved data analysis.
A preliminary set of IRFs have been produced to match the \irf{P7REP} data and
have been labeled \irf{P7REP\_V10}, e.g., \irf{P7REP\_SOURCE\_V10}.
As the various analysis components are still being validated, we
anticipate that the results shown here may change slightly by the time
the \irf{P7REP} data are released. The corresponding IRFs likewise
would be a later iteration (e.g., \irf{P7REP\_V12} or
\irf{P7REP\_V13}) with changes in the effective area at the $\sim
5\%$ level with respect to the \irf{P7REP\_V10} IRFs presented here.
The \Fermi\ Science Support Center
(FSSC)\footnote{\url{http://fermi.gsfc.nasa.gov/ssc/data/}} is 
the authoritative source for recommendations regarding the analysis
of \Fermi\ data.

The \irf{P7REP\_V10} IRFs include our best understanding of the instrument.
As for the \irf{P7\_V6} version available now, they were derived for the event
classes used in standard LAT data analysis and for the tracker thin-converter
section (front) and thick-converter section (back) separately
(see~\citet{REF.2012.P7PerfPaper} for more details). The \irf{P7REP\_V10} IRFs
have been derived as usual using GEANT4-based~\citep{REF:GEANT} MC simulations.
However, both the effective area and the PSF have been refined using
information from flight data.

For the effective area, analysis of photons from the Vela pulsar and the
Galactic Ridge have shown that, when analyzed separately, front- and
back-converting events give inconsistent results for the absolute measured
flux. The observed discrepancy is very likely due to differences between the
true and simulated instrument response and has been a long-standing issue
observed since launch.  Because we cannot determine from flight data whether
the front or back response is more accurate, we applied a symmetric correction
factor to both in order to keep the total ${\rm front} + {\rm back}$ effective
area unchanged. The front/back ratio discrepancy ranges from $-8\%$ at 100~MeV
to $+4\%$ at 300~MeV and greater.

For the PSF, we analyzed \irf{P7REP\_CLEAN} event-class data from the Vela
pulsar and a sample of 40 bright active Galactic nuclei (AGN) and found a
significant improvement in pointing resolution above 1~GeV with respect to
non-reprocessed data. Because the on-orbit PSF is now much closer to the
MC-derived PSF, we developed a new method to parameterize the former.
We have generated an on-orbit PSF model for \irf{P7REP} using the same
King-function parameter tables as the MC-derived PSF \irf{P7CLEAN\_V6MC} but
with different coefficients for the PSF scaling function, which are fit to
match the PSF distributions of the Vela pulsar and AGN calibration data sets.
We find that the PSF of \irf{P7REP\_CLEAN\_V10} is significantly improved
relative to \irf{P7} with a 30\% (40\%) reduction in the PSF 68\% containment
radius for front (back) events above 10~GeV, as shown in figure~\ref{fig:psf68}.
A statistically significant \makebox{20--25\%} residual discrepancy with
respect to \irf{P7CLEAN\_V6MC} remains above 10~GeV for both front and back
events. Overall, we find that the new on-orbit PSF provides a good
representation of the angular dispersion while preserving the dependence on the
\gammaRayHyph\ incidence angle.

\irf{P7REP\_V10} IRF tables are defined up to 1.8~TeV. However the energy
reconstruction has been tested only up to 1~TeV. Therefore the FSSC
data server will only release data up to 1~TeV by default.
Users will be able to change the default query settings to access the events
with energies above 1~TeV.

\section{Validation and performance}

Reprocessing data with up-to-date calibration constants provides us with a new
data set that is significantly different from the one that we explored in the
careful analysis published in~\cite{REF.2012.P7PerfPaper}.
As soon the first \irf{P7REP} data was available we started the low level
verification (e.g., the characterization of the change in the energy scale).
The validation process is still on-going at the highest level of science
analysis, such as a full sky catalog-like analysis. We report here a few
highlights.

\subsection{Events switching event class}

\begin{figure}[htbp!]
  \includegraphics[width=\linewidth]{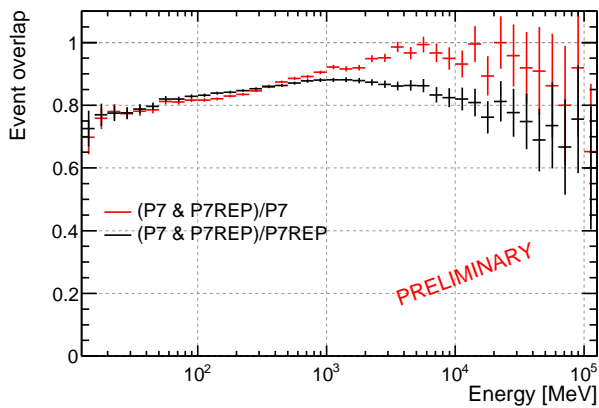}
  \caption{Relative overlap between \irf{P7REP\_SOURCE} and \irf{P7SOURCE}.
    Up to 25\% of the events move from one class to another.}
  \label{fig:evtsclass}
\end{figure}

A first quantification of how reprocessed data are different is given by the
fact that 25\% of the events move from one class to another, as shown in
figure~\ref{fig:evtsclass} for the Source event class. This number was not
unexpected, given the magnitude of the change in the CAL crystal light
asymmetry calibration and its known impact on the tracking.

\subsection{Vela spectral analysis}

% https://confluence.slac.stanford.edu/display/SCIGRPS/2012/06/21/Vela+2+paper+spectral+analysis+with+reprocessed+P7+data
Vela is useful as a reference because of its brightness, which makes the impact
of the diffuse \gammaRayHyph\ backgrounds very limited on most types of
analysis. Furthermore, we can use \gammaRays\ away from the pulse peaks to very
effectively model the diffuse \gammaRayHyph\ backgrounds.

We analyzed both reprocessed and original data using the same procedure, which
may be simply described as follows: we first fit one year of data in a large
region around Vela, leaving all the parameters of the sources in the model free
to vary. We then developed a model for the non-pulsed background near Vela
using the sample of photons in a $10^{\circ}$ region around Vela that are also
away from the Vela phase peaks and that we fit using as input the parameters
for the larger region. We then removed as spurious any sources that were not in
the first Fermi source catalog~\cite{REF:2010.1FGL}. We used this model as the
background template for our phase-averaged analysis.

\begin{figure}[htbp!]
  \includegraphics[width=\linewidth]{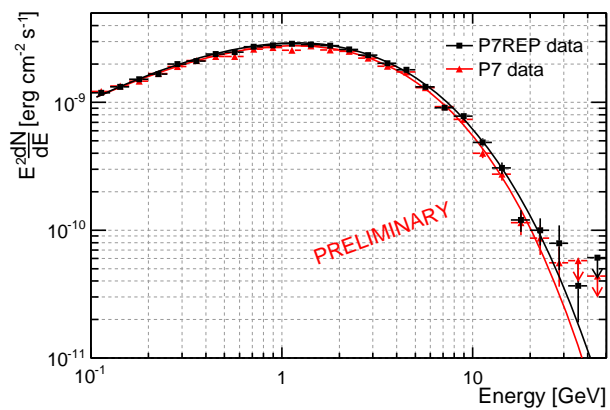}
  \caption{Phase-averaged spectrum of the Vela pulsar, measured with source
    class events using the first year of \irf{P7SOURCE} and \irf{P7REP\_SOURCE}
    data.}
  \label{fig:velaspec}
\end{figure}

The model used to fit the Vela pulsar is a power-law with an exponential cutoff
and is shown for both data sets in figure~\ref{fig:velaspec}. Overall the
results are roughly compatible in the sense that high level science is not
radically changed, but statistically the reconstructed flux for
\irf{P7REP\_SOURCE} data is higher by a few percent and the spectral index is
slightly harder, confirming the impact of the reprocessing on the measurements
performed with the LAT.

\subsection{Geminga flux stability}

\begin{figure}[htbp!]
  \includegraphics[width=\linewidth]{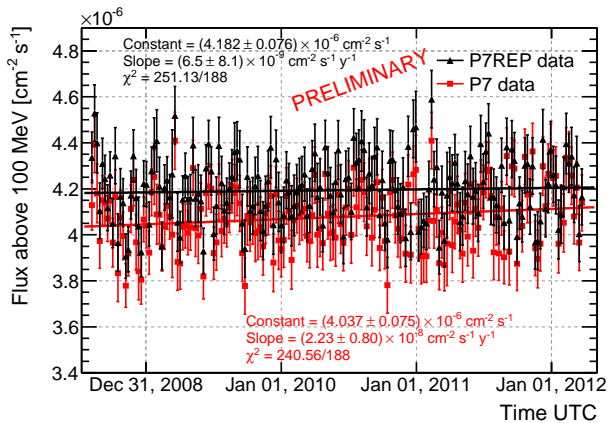}
  \caption{Long-term trending of the integral flux above 100~MeV for Geminga,
    measured using \irf{P7SOURCE} and \irf{P7REP\_SOURCE} class events. Each
    data point represents a week's worth of data.}
  \label{geminga}
\end{figure}

% https://confluence.slac.stanford.edu/display/SCIGRPS/LAT+stability+IRF+testbench
In order to check the stability of LAT data quality and data-analysis chain, we
evaluated weekly binned light curves of the Vela and Geminga pulsars using a
standard likelihood analysis. The resulting light curves are shown in
figure~\ref{geminga} for the Geminga pulsar for 4 years of \irf{P7SOURCE}
and \irf{P7REP\_SOURCE} data. The flux in the \irf{P7REP\_SOURCE} data is again
slightly higher ($\sim 2.7\%$) for both pulsars. In addition there is an
increase in flux by $\sim 0.7\%$ per year detected for Geminga in the
\irf{P7SOURCE} data, which is not detected in the \irf{P7REP\_SOURCE} data
for which the flux is stable within uncertainties over 4 years.

This result demonstrates that \irf{P7REP} data are of better quality than
\irf{P7} data, and have lower systematic uncertainties.

\subsection{Hard-spectrum source list study}

% https://confluence.slac.stanford.edu/display/SCIGRPS/2012/06/27/Hard+sources+with+reprocessed+P7+data
A preliminary higher-level test was provided by a comprehensive (re)analysis of
$\sim 1300$ candidate hard sources that had a test statistics (TS) $>10$ for 3
years of \irf{P7} data. We choose the harder sources for this analysis as they
were more likely to be affected by improvements to the PSF at high energies.
Again, both \irf{P7CLEAN} and \irf{P7REP\_CLEAN} data sets between 10 and
500~GeV were analyzed in parallel. We note that the \irf{P7REP\_CLEAN} analysis
used a preliminary version of the isotropic spectral template that was derived
from the \irf{P7REP} data.

\begin{figure}[htb!]
  \includegraphics[width=\linewidth]{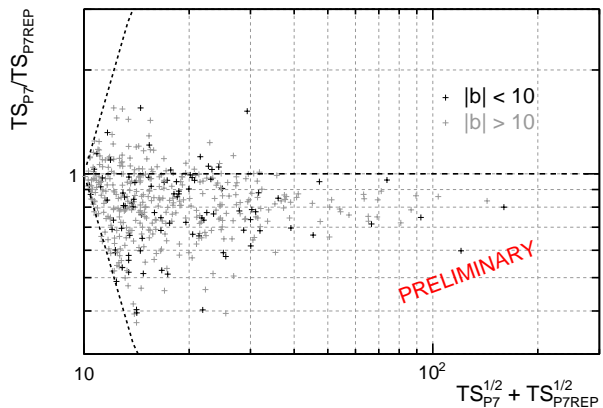}
  \caption{Ratio between the TS values obtained from the analysis of
    \irf{P7CLEAN}- and \irf{P7REP\_CLEAN}-class events for a sample of candidate
    hard sources as a function of the sum of the detection significances.}
  \label{fig:hslTS}
\end{figure}

There are 514 sources at ${\rm TS} >25$ in the \irf{P7CLEAN} run, 561 with the
\irf{P7REP\_CLEAN} data and 454 are in common. Again the relatively large
number of sources not in common is to be expected, since \irf{P7REP} data are
really a new data set, and most of the sources in both lists are close to the
TS threshold. Therefore small changes in TS can cause substantial changes in
the source lists. The TS distributions for both data analysis runs are
presented in figure~\ref{fig:hslTS}. Overall the \irf{P7REP\_CLEAN} data give a
slightly higher TS: on average the ratio \irf{P7CLEAN}/\irf{P7REP\_CLEAN}
is 0.82.

\section{Conclusions}

% https://confluence.slac.stanford.edu/display/SCIGRPS/2013/02/11/Proposed+text+regarding+P7+reprocessing
We have reprocessed \fermi-LAT data with up-to-date calibration constants,
including a more accurate description of the position-dependent response of
each CAL scintillator crystal and the slight decrease in scintillation light
yield with time ($\sim 1\%$ per year) from radiation exposure on orbit.
The main improvement to the instrument response is a narrower PSF core above a
few GeV. For this reason, we have produced a new set of IRFs that include the
new on-orbit PSF which is now based on a scaled version of the PSF derived from
Monte Carlo simulations. The new IRFs also include a modification of the
effective area, inferred from flux measurements using flight data, to correct
symmetrically the front/back discrepancy in the effective area observed in
the data.

Validation of the reprocessed data is ongoing within the collaboration and has
already demonstrated an overall improvement in data quality, which has led to
better standard source analysis. We expect to release these new data publicly
through the FSSC in the spring of 2013.
The main results shown in this paper will be updated to the final versions
of the \irf{P7REP} IRFs and made available through the FSSC as well.

\bigskip % extra skip inserted

\begin{acknowledgments}
  The $Fermi$ LAT Collaboration acknowledges support from a number of agencies
  and institutes for both development and the operation of the LAT as well as
  scientific data analysis. These include NASA and DOE in the United States,
  CEA/Irfu and IN2P3/CNRS in France, ASI and INFN in Italy, MEXT, KEK, and JAXA
  in Japan, and the K.~A.~Wallenberg Foundation, the Swedish Research Council
  and the National Space Board in Sweden. Additional support from INAF in Italy
  and CNES in France for science analysis during the operations phase is also
  gratefully acknowledged. 
\end{acknowledgments}

\bigskip % extra skip inserted

\bibliographystyle{apsrev}
\bibliography{p202proc}

\begin{thebibliography}{4}
\expandafter\ifx\csname natexlab\endcsname\relax\def\natexlab#1{#1}\fi
\expandafter\ifx\csname bibnamefont\endcsname\relax
  \def\bibnamefont#1{#1}\fi
\expandafter\ifx\csname bibfnamefont\endcsname\relax
  \def\bibfnamefont#1{#1}\fi
\expandafter\ifx\csname citenamefont\endcsname\relax
  \def\citenamefont#1{#1}\fi
\expandafter\ifx\csname url\endcsname\relax
  \def\url#1{\texttt{#1}}\fi
\expandafter\ifx\csname urlprefix\endcsname\relax\def\urlprefix{URL }\fi
\providecommand{\bibinfo}[2]{#2}
\providecommand{\eprint}[2][]{\url{#2}}

\bibitem[{\citenamefont{{Abdo} et~al.}(2009)\citenamefont{{Abdo}, {Ackermann},
  {Ajello}, {Ampe}, {Anderson}, {Atwood}, {Axelsson}, {Bagagli}, {Baldini},
  {Ballet} et~al.}}]{REF.2009.OnOrbitCalib}
\bibinfo{author}{\bibfnamefont{A.~A.} \bibnamefont{{Abdo}}},
  \bibinfo{author}{\bibfnamefont{M.}~\bibnamefont{{Ackermann}}},
  \bibinfo{author}{\bibfnamefont{M.}~\bibnamefont{{Ajello}}},
  \bibinfo{author}{\bibfnamefont{J.}~\bibnamefont{{Ampe}}},
  \bibinfo{author}{\bibfnamefont{B.}~\bibnamefont{{Anderson}}},
  \bibinfo{author}{\bibfnamefont{W.~B.} \bibnamefont{{Atwood}}},
  \bibinfo{author}{\bibfnamefont{M.}~\bibnamefont{{Axelsson}}},
  \bibinfo{author}{\bibfnamefont{R.}~\bibnamefont{{Bagagli}}},
  \bibinfo{author}{\bibfnamefont{L.}~\bibnamefont{{Baldini}}},
  \bibinfo{author}{\bibfnamefont{J.}~\bibnamefont{{Ballet}}},
  \bibnamefont{et~al.}, \bibinfo{journal}{Astroparticle Physics}
  \textbf{\bibinfo{volume}{32}}, \bibinfo{pages}{193} (\bibinfo{year}{2009}),
  \eprint{0904.2226}.

\bibitem[{\citenamefont{{Ackermann} et~al.}(2012)\citenamefont{{Ackermann},
  {Ajello}, {Albert}, {Allafort}, {Atwood}, {Axelsson}, {Baldini}, {Ballet},
  {Barbiellini}, {Bastieri} et~al.}}]{REF.2012.P7PerfPaper}
\bibinfo{author}{\bibfnamefont{M.}~\bibnamefont{{Ackermann}}},
  \bibinfo{author}{\bibfnamefont{M.}~\bibnamefont{{Ajello}}},
  \bibinfo{author}{\bibfnamefont{A.}~\bibnamefont{{Albert}}},
  \bibinfo{author}{\bibfnamefont{A.}~\bibnamefont{{Allafort}}},
  \bibinfo{author}{\bibfnamefont{W.~B.} \bibnamefont{{Atwood}}},
  \bibinfo{author}{\bibfnamefont{M.}~\bibnamefont{{Axelsson}}},
  \bibinfo{author}{\bibfnamefont{L.}~\bibnamefont{{Baldini}}},
  \bibinfo{author}{\bibfnamefont{J.}~\bibnamefont{{Ballet}}},
  \bibinfo{author}{\bibfnamefont{G.}~\bibnamefont{{Barbiellini}}},
  \bibinfo{author}{\bibfnamefont{D.}~\bibnamefont{{Bastieri}}},
  \bibnamefont{et~al.}, \bibinfo{journal}{\apjs}
  \textbf{\bibinfo{volume}{203}}, \bibinfo{eid}{4} (\bibinfo{year}{2012}),
  \eprint{1206.1896}.

\bibitem[{\citenamefont{{Agostinelli} et~al.}(2003)}]{REF:GEANT}
\bibinfo{author}{\bibfnamefont{S.}~\bibnamefont{{Agostinelli}}}
  \bibnamefont{et~al.} (\bibinfo{collaboration}{GEANT4}),
  \bibinfo{journal}{Nucl. Instrum. Meth.} \textbf{\bibinfo{volume}{A506}},
  \bibinfo{pages}{250} (\bibinfo{year}{2003}).

\bibitem[{\citenamefont{{Abdo} et~al.}(2010)\citenamefont{{Abdo}, {Ackermann},
  {Ajello}, {Allafort}, {Antolini}, {Atwood}, {Axelsson}, {Baldini}, {Ballet},
  {Barbiellini} et~al.}}]{REF:2010.1FGL}
\bibinfo{author}{\bibfnamefont{A.~A.} \bibnamefont{{Abdo}}},
  \bibinfo{author}{\bibfnamefont{M.}~\bibnamefont{{Ackermann}}},
  \bibinfo{author}{\bibfnamefont{M.}~\bibnamefont{{Ajello}}},
  \bibinfo{author}{\bibfnamefont{A.}~\bibnamefont{{Allafort}}},
  \bibinfo{author}{\bibfnamefont{E.}~\bibnamefont{{Antolini}}},
  \bibinfo{author}{\bibfnamefont{W.~B.} \bibnamefont{{Atwood}}},
  \bibinfo{author}{\bibfnamefont{M.}~\bibnamefont{{Axelsson}}},
  \bibinfo{author}{\bibfnamefont{L.}~\bibnamefont{{Baldini}}},
  \bibinfo{author}{\bibfnamefont{J.}~\bibnamefont{{Ballet}}},
  \bibinfo{author}{\bibfnamefont{G.}~\bibnamefont{{Barbiellini}}},
  \bibnamefont{et~al.}, \bibinfo{journal}{\apjs}
  \textbf{\bibinfo{volume}{188}}, \bibinfo{pages}{405} (\bibinfo{year}{2010}),
  \eprint{1002.2280}.

\end{thebibliography}

\end{document}